\documentclass[10pt,twocolumn,a4paper]{article}
\usepackage[T1]{fontenc}
\usepackage[utf8]{inputenc}
\usepackage{mathptmx}
\usepackage{courier}
\usepackage[margin=1.8cm,columnsep=0.6cm]{geometry}
\usepackage{amsmath,amssymb}
\usepackage{graphicx}
\usepackage{booktabs}
\usepackage[authoryear,round]{natbib}
\usepackage[font=small,labelfont=bf]{caption}
\usepackage{xcolor}
\usepackage{xurl}
\usepackage[colorlinks=true,allcolors=blue!60!black]{hyperref}

\title{\bfseries Sky Monitor AI: classifying night-sky conditions from an observatory all-sky camera with a CNN ensemble, label-free sky types and exact evidence maps}
\author{Muhammad Waqas\\[2pt]
\small Department of Astronomy, Institut Teknologi Bandung, Jl.\ Ganesha 10, Bandung 40132, Indonesia\\
\small \href{mailto:mdwaqas767@gmail.com}{mdwaqas767@gmail.com}}
\date{}

\begin{document}
\twocolumn[
\maketitle
\begin{abstract}
Small and remotely operated observatories must decide every night, often without staff on site, whether the sky is clear enough to observe. We present a complete pipeline that classifies frames of the auxiliary all-sky camera of the Higashi-Hiroshima Observatory as clear or cloudy, evaluated on 8\,806 frames from 32 nights. A circular sky mask removes a printed date stamp that would otherwise act as a shortcut, and all evaluations use whole held-out nights. Hand-crafted sky statistics reach 77.1 per cent accuracy on six test nights; a small convolutional neural network (CNN) trained from scratch reaches 84.3 per cent, and an ensemble of the small CNN and two ImageNet-pretrained ResNet-18 variants, selected on validation nights, reaches 86.7 per cent (AUC 0.924) with the smallest night-to-night scatter (0.879\,$\pm$\,0.086). Ephemeris calculations show that moonlight is the dominant difficulty: across Moon altitude the ensemble's accuracy varies by 5 percentage points, the hand-crafted model's by 20. Fine-tuning deeper ResNet layers overfits with $\sim$24 training nights, which underlines the need to match model capacity to the number of independent nights. Following the labels through the night shows that they were assigned in blocks of 1--2\,h, and that a large share of the remaining errors are mislabelled or partly cloudy frames. Without labels, Gaussian-mixture clustering of pretrained CNN embeddings recovers the clear/cloudy split with 87 per cent purity and isolates frames with water droplets on the camera dome, a class absent from the labels. Gradient-based maps and a quantitative rim test indicate that the networks rely on sky regions rather than on fixed structures. Finally, we show that for networks ending in global average pooling the decision decomposes exactly into a map of local logits, which yields gradient-free evidence maps and a cloud-cover estimate, and we release an open-source application, Sky Monitor AI, that applies the models to single frames, whole nights and live cameras.
\end{abstract}
\par\vspace{4pt}
\vspace{10pt}
]

\section{Introduction}
\label{sec:intro}

Clouds are the main reason an optical telescope loses observing time, and the weather that accompanies them (humidity, dew and rain) is a threat to optics and mechanics. Every observatory therefore needs a rule for when to open the dome. At staffed sites an observer makes the call; at small university observatories, and at robotic or remotely operated facilities, the decision must increasingly be taken from instruments. All-sky cameras, inexpensive cameras with fisheye lenses that image the whole sky at regular intervals, are a natural sensor for this task: a human can judge the sky from a single frame at a glance.

Automating that glance has a long history. \citet{Shamir2005} mapped relative opacity from night-time panoramic images using the visibility of catalogued stars. \citet{Mommert2020} classified sub-regions of all-sky images at Lowell Observatory with gradient-boosted trees on hand-designed features (star density, sky brightness and temporal changes), reaching about 95 per cent accuracy and outperforming a ResNet model in that setting. More recently, \citet{Haghighi2025} applied deep learning to identify precipitation clouds for observatory safety, and \citet{Buntin2025} detected, tracked and forecast clouds by differencing successive frames. At the Higashi-Hiroshima Observatory itself, the sky-condition model used operationally is described in the appendix of \citet{Uemura2025}, in the context of an autonomous follow-up observation framework.

This paper presents a complete study for one camera, the auxiliary all-sky camera of the Higashi-Hiroshima Observatory, on a dataset distributed as a project of the AstroAI Network Summer School (2026). Rather than maximising a single score, we focus on the questions an observatory needs answered before trusting such a system: does the model use the sky or a shortcut; how well does it generalise to nights it has never seen; what limits its accuracy; can the structure of the data be found without labels; and how can the model's reasoning be shown to an operator? Our contributions are:
\begin{enumerate}
\setlength{\itemsep}{1pt}
\item a night-based evaluation of hand-crafted features, a small CNN and ImageNet-pretrained ResNet-18 models, including a validation-selected ensemble (Sections~\ref{sec:methods}--\ref{sec:results});
\item a quantitative analysis of the effect of moonlight, using the Moon's altitude computed for every frame, and of label quality, by following labels and predictions through whole nights (Section~\ref{sec:results});
\item label-free discovery of sky types from pretrained CNN embeddings, including a wet-dome class (Section~\ref{sec:unsup});
\item an interpretation study with signed Grad-CAM, a quantitative rim test and blur-based occlusion (Section~\ref{sec:interp});
\item an exact, gradient-free decomposition of the decision of global-average-pooling networks into local logits, used for evidence maps and cloud-cover estimates in an open-source tool (Section~\ref{sec:tool}).
\end{enumerate}

\section{Data}
\label{sec:data}

\subsection{Camera and frames}
\begin{figure*}
\centering
\includegraphics[width=0.82\textwidth]{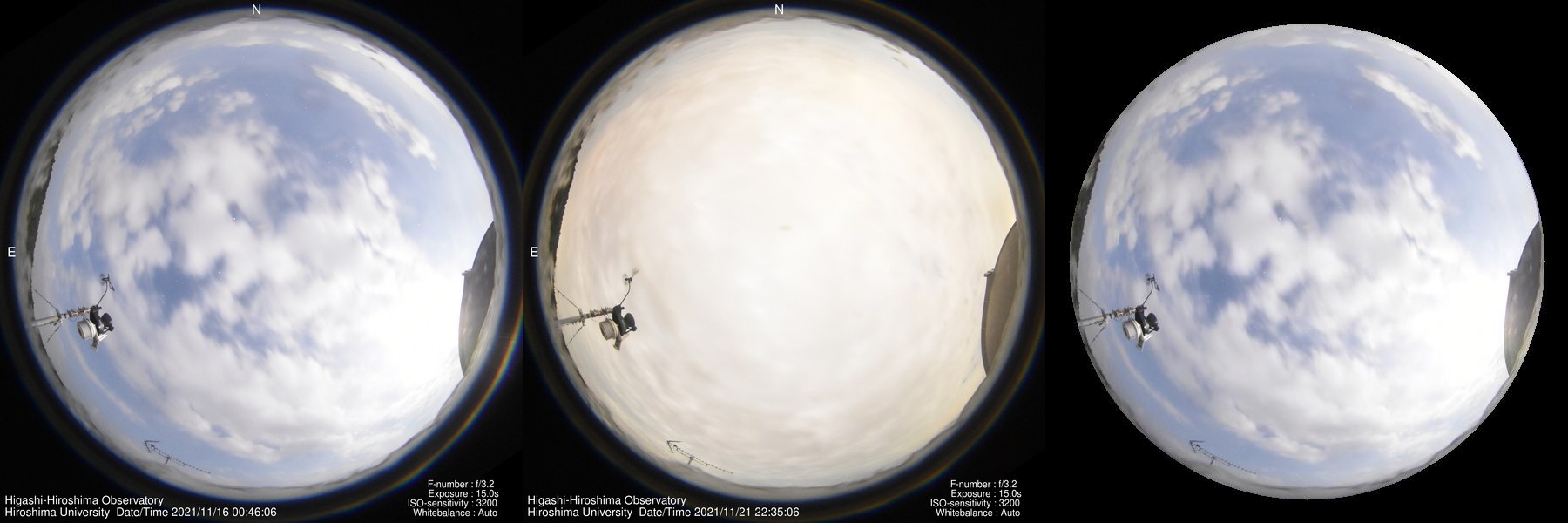}
\caption{Frames from the dataset. Left: a frame labelled \emph{fine}, taken in moonlight at 00:46 local time on 2021 November 16; broken cloud is visible despite the label. Centre: a frame labelled \emph{cloudy} (2021 November 21), a bright overcast sky. Right: the first frame after applying the circular sky mask of radius $0.455\,W$, which removes the printed caption, including the date.}
\label{fig:frames}
\end{figure*}

The camera records $600\times600$ pixel colour frames through a fisheye lens, mapping the visible hemisphere onto a circle with north at the top and east on the left. A caption printed outside the sky circle records the exposure settings (f/3.2, 15\,s, ISO\,3200, automatic white balance), the observatory name and the date and local time. Two fixed structures intrude into the field of view: a mast on the left and the edge of a building on the right. At night three light sources dominate: stars, which are the most direct evidence of a clear sky; the Moon, whose light scattered by a clear atmosphere makes the sky appear blue in a 15-s exposure; and artificial light from the town, which illuminates the base of low clouds in orange (Fig.~\ref{fig:frames}).

\subsection{Dataset and labels}
The frames were provided pre-sorted into two folders, \emph{fine} and \emph{cloudy}. The working set contains 8\,806 frames, 4\,984 cloudy and 3\,822 fine (57 per cent cloudy); the fine folder includes a small number of duplicated files, which share their night and therefore cannot leak between data splits. File names are timestamps (\texttt{YYMMDDhhmmss}) in Japan Standard Time (UTC+9), which we verified against the printed captions. Defining an observing night by shifting each timestamp back by 12\,h, the data span 32 nights between 2021 November 15 and December 19. Fine frames come from 12 nights and cloudy frames from 20; four nights contain both classes. The camera takes a frame roughly once per minute, so consecutive frames are nearly identical and the number of effectively independent samples is closer to the number of nights than to the number of frames.

Two properties of the data shape the analysis. First, the classes are partly separated in time: all fine frames were taken before November 30, while cloudy frames continue to December 19. Any date-dependent property (the printed date, the lunar phase) is therefore correlated with the label. Second, the full Moon of 2021 November 19 falls in the middle of a run of clear nights, so bright clear skies in the data are mostly moonlit.

\subsection{Sky mask}
\label{sec:mask}
Because the printed date would allow a network to predict the label without looking at the sky, a form of shortcut learning \citep{Geirhos2020}, every pixel outside a centred circle of radius $r=0.455\,W$ ($\approx273$ pixels) is set to zero in every experiment. The mask removes the captions and margins and keeps the whole sky; the mast and building edge lie inside the circle and remain. The masked fraction of the image is $f = 1-\pi\,0.455^2 \approx 0.35$. A centred circular mask is invariant under rotations and flips, which allows these as augmentations.

The mask has a subtle consequence for image statistics. If a fraction $f$ of the pixels is zero and the sky pixels have mean $b$ and standard deviation $\sigma_{\rm sky}$, the whole-image mean and variance are $m=(1-f)\,b$ and $\sigma^2=(1-f)\,\sigma_{\rm sky}^2+f(1-f)\,b^2$, so that
\begin{equation}
\sigma \;\geq\; m\,\sqrt{f/(1-f)} \approx 0.73\,m .
\end{equation}
A scatter plot of whole-image standard deviation against mean therefore collapses onto a line of slope $\approx0.73$, a property of the mask geometry rather than of the sky. All statistics below are computed over sky pixels only.

\subsection{Night-based split}
All frames of a night are assigned to the same subset (Table~\ref{tab:split}). Nights were assigned separately per class according to the dominant class of each night, so that every subset contains both classes. The mixed night of November 24 (298 cloudy, 240 fine frames), on which Moon, season and camera settings are identical for both classes, was placed in the test set. The test set was used only for the final evaluation; all model choices were made on the validation nights. Classes were weighted in the loss as $w_c = N/(K N_c)$ with $K=2$, giving $w_{\rm cloudy}=0.91$ and $w_{\rm fine}=1.12$.

\begin{table}
\centering
\caption{Split of the data by observing night.}
\label{tab:split}
\small
\begin{tabular}{@{}lrrr@{}}
\toprule
Subset & Frames & Cloudy & Fine\\
\midrule
Training (21 nights) & 5\,705 & 3\,150 & 2\,555\\
Validation (5 nights) & 1\,495 & 871 & 624\\
Test (6 nights) & 1\,606 & 963 & 643\\
\bottomrule
\end{tabular}

\vspace{2pt}
\raggedright\footnotesize Validation: 2021 Nov 20, Nov 22, Dec 3, Dec 7, Dec 9. Test: Nov 15, Nov 24, Nov 29, Dec 11, Dec 15, Dec 19.
\end{table}

\section{Methods}
\label{sec:methods}

\subsection{Hand-crafted baseline}
\label{sec:baseline}
\begin{figure*}
\centering
\includegraphics[width=0.92\textwidth]{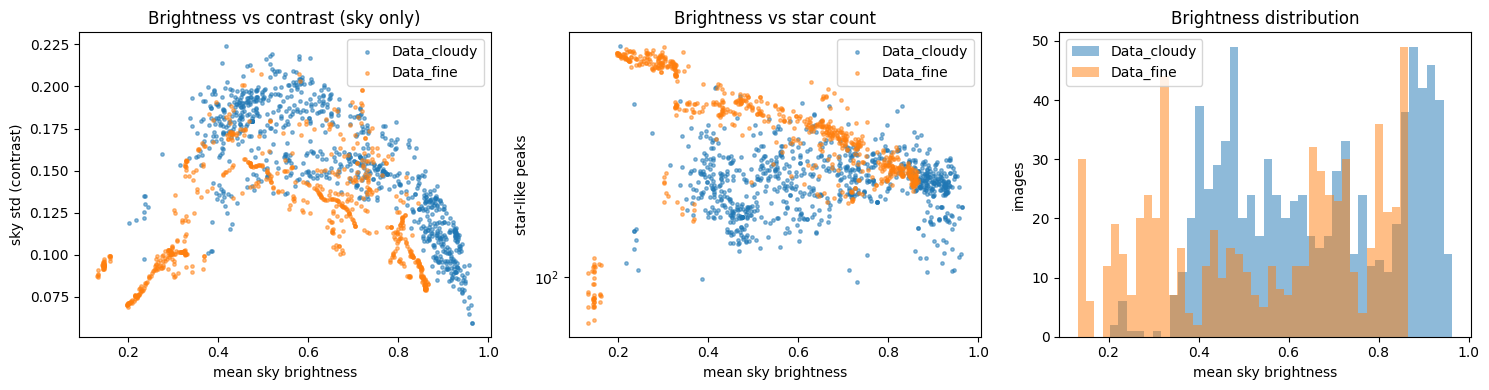}
\caption{Sky-only statistics for 1\,500 frames. Left: mean sky brightness against contrast. Centre: brightness against the number of star-like peaks (logarithmic axis). Right: brightness distribution per class. Fine frames form continuous tracks (individual clear nights seen minute by minute); cloudy frames scatter. Above a brightness of $\sim$0.8 the star-count bands merge because the bright sky washes out the stars.}
\label{fig:features}
\end{figure*}

Four physically motivated features were computed on the grey-scale sky pixels: the mean brightness; the standard deviation (contrast); the mean gradient magnitude after light smoothing (texture); and the number of star-like peaks, defined as pixels that are the maximum of their $5\times5$ neighbourhood and exceed a $9\times9$ local mean by more than 0.06 on a 0--1 scale. On 1\,500 random frames the single-feature AUCs are 0.654, 0.741, 0.775 and 0.810, respectively (Fig.~\ref{fig:features}): brightness is the weakest cue because the Moon brightens clear and cloudy skies alike, and the star count, the astronomer's own test, is the strongest. A gradient-boosting classifier on the four features reaches 84.5 per cent accuracy (AUC 0.870) in five-fold cross-validation grouped by night, against 82.2 per cent for logistic regression and 56.9 per cent for the majority class. For comparison with the networks, the gradient-boosting model was retrained on 2\,000 training frames and evaluated on the test nights.

\subsection{Convolutional networks}
Masked frames were resized to $224\times224$ pixels with bilinear interpolation, scaled to $[0,1]$ and normalised with the ImageNet channel means and standard deviations. Training used the weighted cross-entropy loss, the AdamW optimiser \citep{Loshchilov2019} with weight decay $10^{-4}$, mini-batches of 32, a cosine learning-rate schedule, and early stopping on the validation loss. Training augmentations were horizontal and vertical flips, rotations by up to $180^\circ$, and brightness and contrast changes. All models were implemented in PyTorch \citep{Paszke2019} and trained on a single NVIDIA T4 GPU.

\paragraph{Small CNN.} Four blocks of $3\times3$ convolution, batch normalisation \citep{Ioffe2015}, ReLU and $2\times2$ max pooling with 32, 64, 128 and 256 filters, followed by global average pooling, dropout (0.3) and a linear layer with two outputs; 389\,890 parameters in total. It was trained for at most 25 epochs from a learning rate of $10^{-3}$ with patience 5; training stopped after 15 epochs and the epoch with the lowest validation loss (epoch 10) was kept.

\paragraph{ResNet-18.} ResNet-18 \citep{He2016} pretrained on ImageNet \citep{Deng2009} (11.2 million parameters), with its classifier replaced by a two-class linear layer. A first run trained the head with the backbone frozen and then fine-tuned all layers at a learning rate of $10^{-4}$; as soon as all layers were unfrozen the training accuracy rose to 0.99 while the validation loss rose from $\approx$0.3 to 0.82--1.55, i.e.\ the network memorised the training nights. In the second configuration (\emph{v2}) only the last stage (layer4, $\sim$8.4 million parameters) was unfrozen in stage 2, at a learning rate of $3\times10^{-5}$, with stronger augmentation (brightness $\pm$50 per cent, contrast $\pm$30 per cent), and the best model was kept across both stages. Stage 2 again increased the validation loss (to 0.53--0.82), so the selected v2 model is the stage-1 model: a frozen ImageNet backbone with a trained linear head (validation loss 0.297, accuracy 0.906). The third configuration (\emph{v3}) added five safeguards: batch-normalisation layers kept in evaluation mode with their ImageNet statistics; an L2-SP penalty \citep{Li2018} $\alpha\sum_{i\in{\rm layer4}}(w_i-w_i^{\rm pre})^2$ with $\alpha=0.01$; label smoothing \citep{Szegedy2016} with $\varepsilon=0.1$; physically motivated augmentation (random gamma 0.6--1.6, Gaussian blur on 30 per cent of images to mimic dew or haze, Gaussian noise of standard deviation 0.02, and random erasing on 30 per cent of images to mimic obstructions); and a learning rate of $10^{-5}$ with patience 3. With these safeguards the validation loss remained between 0.31 and 0.35 during fine-tuning, but the best value (0.309, accuracy 0.892) did not improve on v2.

\subsection{Ensemble and model selection}
The final model averages the fine-class probabilities of the $M=3$ networks, $P_{\rm ens} = M^{-1}\sum_m P_m$. If each model's error has variance $\sigma^2$ and pairwise correlation $\rho$, the variance of the averaged error is $\rho\sigma^2+(1-\rho)\sigma^2/M$, so averaging helps when the models fail on different inputs. The decision to use an ensemble, and the choice among ensembles, were made on the validation nights (Table~\ref{tab:val}): the three-model ensemble had the highest validation accuracy and was selected before the test results were examined in detail.

\begin{table}
\centering
\caption{Validation results used for model selection.}
\label{tab:val}
\small
\begin{tabular}{@{}lcc@{}}
\toprule
Model & AUC & Accuracy\\
\midrule
Small CNN & 0.953 & 0.854\\
ResNet-18 v2 & 0.919 & 0.906\\
ResNet-18 v3 & 0.927 & 0.892\\
Ensemble: small CNN + v2 & \textbf{0.959} & 0.901\\
Ensemble: all three (final) & 0.954 & \textbf{0.908}\\
\bottomrule
\end{tabular}
\end{table}

\subsection{Moon ephemeris and temporal smoothing}
For every frame the Moon's altitude and illuminated fraction were computed with Astropy \citep{Astropy2022} from the UTC time and the approximate site coordinates (34.38$^\circ$\,N, 132.78$^\circ$\,E, $\sim$500\,m), neglecting refraction and approximating the phase angle by the elongation; both simplifications are negligible at the level of binning used here. To exploit the temporal coherence of the sky, we also evaluate a centred rolling median of $P_{\rm ens}$ over windows of $n$ consecutive frames within a night.

\section{Classification results}
\label{sec:results}

\subsection{Held-out nights}
Table~\ref{tab:results} summarises the test results. On the same six nights the hand-crafted model drops from its cross-validated 84.5 per cent to 77.1 per cent, and every network exceeds it by 4--7 percentage points in accuracy and 0.04--0.07 in AUC. The small CNN is the best single model. ResNet-18 v2 reached 0.906 accuracy on the validation nights but 0.812 on the test nights: with five or six nights per subset, \emph{which} nights fall into which subset matters as much as the choice of model. The two ResNet models have similar accuracy but very different error balance: v3, with frozen normalisation statistics, is conservative (66 cloudy frames called fine, against 183 for v2) but misses many clear frames (230 against 119). For equipment safety this conservative behaviour may be preferable.

\begin{table*}
\centering
\caption{Results on the six test nights (1\,606 frames; decision threshold 0.5). The per-night columns give accuracy on each night.}
\label{tab:results}
\small
\setlength{\tabcolsep}{4.5pt}
\begin{tabular}{@{}lcc cccccc cc@{}}
\toprule
& & & \multicolumn{6}{c}{Accuracy per test night (2021)} & & \\
\cmidrule(lr){4-9}
Model & AUC & Accuracy & Nov 15 & Nov 24 & Nov 29 & Dec 11 & Dec 15 & Dec 19 & Mean\,$\pm$\,sd & Worst\\
 & & & fine & mixed & fine & cloudy & cloudy & cloudy & & \\
\midrule
Hand-crafted (grad.\ boosting) & 0.856 & 0.771 & 0.372 & 0.716 & 1.000 & 0.931 & 1.000 & 1.000 & 0.836\,$\pm$\,0.231 & 0.372\\
Small CNN & 0.922 & 0.843 & 0.901 & 0.799 & 1.000 & 0.843 & 0.697 & 0.812 & 0.842\,$\pm$\,0.093 & 0.697\\
ResNet-18 v2 & 0.892 & 0.812 & 0.843 & 0.814 & 1.000 & 0.894 & 0.909 & 0.420 & 0.813\,$\pm$\,0.185 & 0.420\\
ResNet-18 v3 & 0.892 & 0.816 & 0.724 & 0.799 & 0.509 & 0.980 & 0.970 & 0.762 & 0.791\,$\pm$\,0.159 & 0.509\\
Ensemble: small CNN + v2 & -- & -- & 0.908 & 0.818 & 1.000 & 0.880 & 0.879 & 0.707 & 0.865\,$\pm$\,0.089 & 0.707\\
\textbf{Ensemble: all three (final)} & \textbf{0.924} & \textbf{0.867} & 0.891 & 0.816 & 1.000 & 0.931 & 0.909 & 0.729 & \textbf{0.879\,$\pm$\,0.086} & \textbf{0.729}\\
\bottomrule
\end{tabular}
\end{table*}

The per-night breakdown shows that the models fail on different nights. The hand-crafted model behaves essentially as a brightness detector: it is perfect on the cloudy nights and on the dark clear night of November 29, but correct for only 37 per cent of the frames of the moonlit clear night of November 15. ResNet-18 v2 fails in the opposite direction on December 19 (0.420), where about 105 of its 183 cloudy-called-fine errors occur. The final ensemble reaches 86.7 per cent accuracy, 2.4 points above the best single model and 9.6 points above the hand-crafted model on the same nights, with the smallest night-to-night scatter; its worst night is still at 0.73.

\subsection{Moonlight}
Table~\ref{tab:moon} groups the test frames by the Moon's altitude. All models perform best with the Moon below the horizon, but they differ strongly in how much they degrade: the hand-crafted model and ResNet-18 v2 lose about 20 percentage points across Moon conditions, the ensemble only 5. The worst bin is 20--40$^\circ$ rather than the zenith. A plausible explanation is that the Moon then lies inside the field but off-centre, creating glare and a brightness gradient that resemble bright cloud on one side of the frame; we regard this as a hypothesis. The only test night on which the Moon never rose (November 29, maximum altitude $-49^\circ$) was classified perfectly by all models except the conservative v3, and the hardest night (December 19) was the night of the full Moon, with the Moon reaching $80^\circ$. With six test nights, altitude bins are partly confounded with individual nights and differ in class balance (29--49 per cent fine), so these trends are strong evidence rather than proof.

\begin{table}
\centering
\caption{Test accuracy grouped by the Moon's altitude.}
\label{tab:moon}
\small
\setlength{\tabcolsep}{4pt}
\begin{tabular}{@{}lrcccc@{}}
\toprule
Moon alt. & $N$ & Hand-cr. & CNN & v2 & Ens.\\
\midrule
$<0^\circ$ & 523 & 0.851 & 0.860 & 0.901 & 0.889\\
0--20$^\circ$ & 201 & 0.736 & 0.866 & 0.771 & 0.861\\
20--40$^\circ$ & 362 & 0.649 & 0.751 & 0.715 & 0.837\\
$>40^\circ$ & 520 & 0.788 & 0.881 & 0.806 & 0.867\\
\midrule
Range & & 0.20 & 0.13 & 0.19 & \textbf{0.05}\\
\bottomrule
\end{tabular}
\end{table}

\subsection{Temporal smoothing and label quality}
A centred rolling median of the ensemble probability raises the test accuracy from 0.867 to 0.872, 0.872, 0.877 and 0.878 for windows of 3, 5, 9 and 15 frames. The gain is small and grows slowly with the window, which indicates that the errors come in long runs rather than as isolated frames: they are systematic. Of the ensemble's 214 errors, 99 occur on November 24, 49 on December 19, 32 on November 15, 31 on December 11 and 3 on December 15.

\begin{figure*}
\centering
\includegraphics[width=0.86\textwidth]{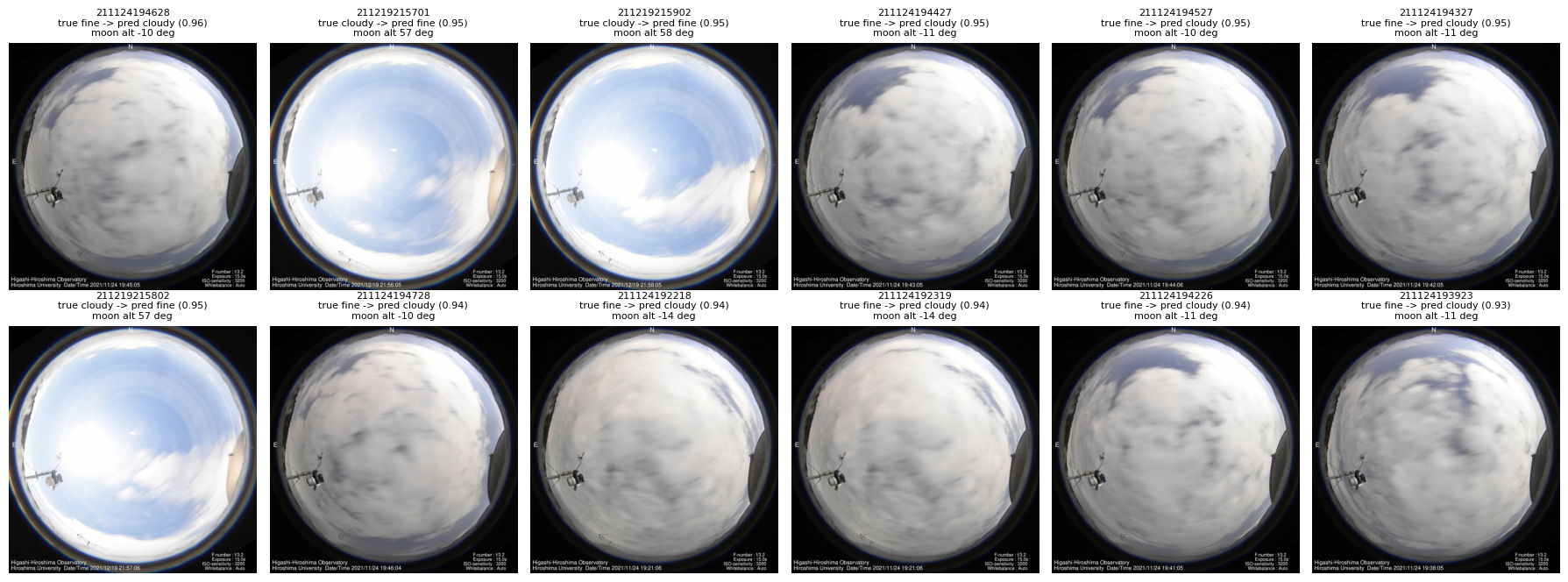}
\caption{The twelve most confident errors of the final ensemble. Nine are from 2021 November 24, 19:21--19:46, labelled fine and predicted cloudy with 93--96 per cent confidence, with the Moon below the horizon; the sky is overcast. Three are from December 19 near 21:57, labelled cloudy and predicted clear with 95 per cent confidence under a full Moon at 57$^\circ$; most of the sky is clear, with cloud near the horizon.}
\label{fig:errors}
\end{figure*}

\begin{figure*}
\centering
\includegraphics[width=0.78\textwidth]{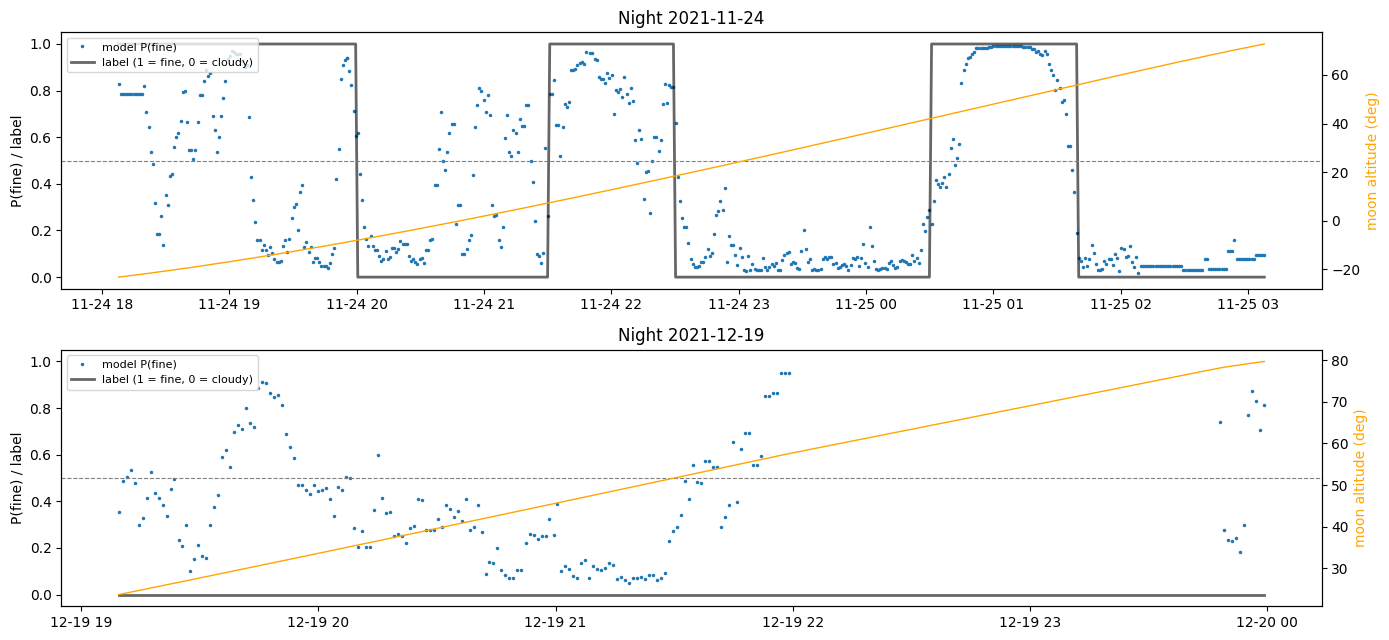}
\caption{Ensemble $P({\rm clear})$ for every frame (dots) of the nights of 2021 November 24 (top) and December 19 (bottom), with the dataset label (black line; 1 = fine) and the Moon's altitude (orange, right axis). The labels form blocks 1--2\,h long, while the model follows changes on a scale of minutes.}
\label{fig:timeline}
\end{figure*}

Inspecting the most confident errors (Fig.~\ref{fig:errors}) and following predictions and labels through whole nights (Fig.~\ref{fig:timeline}) explains this pattern. The labels form rectangular blocks 1--2\,h long that switch at round-number times: time ranges, not individual frames, were labelled. Where the sky is steady, the model agrees with the labels almost perfectly. On November 24, however, the model assigns $P({\rm clear})\approx0.05$--0.2 for about an hour (19:00--19:55) inside a block labelled fine; the corresponding frames show an overcast sky lit from below by the town, with the Moon below the horizon, so the label rather than the model appears to be wrong. Within a block labelled cloudy (20:00--21:30) the model alternates between 0.1 and 0.8, consistent with clouds coming and going. The night of December 19 is labelled cloudy as a single block, but the model shows three excursions to 0.8--0.95, including the bright, mostly clear frames under the full Moon in Fig.~\ref{fig:errors}; these look like genuine breaks in the cloud. The measured accuracy therefore probably underestimates the model, and binary labels cannot describe partly cloudy skies, for which intermediate probabilities are the appropriate response.

\section{Label-free sky types}
\label{sec:unsup}

To test whether the data separate into clear and cloudy skies without labels, 4\,000 random frames were described in two ways: by 21 hand-crafted sky statistics (a 16-bin grey-level histogram, mean brightness, contrast, mean gradient, the logarithm of the star-peak count and the spread between the 99th and 50th brightness percentiles), and by the 512-dimensional global-average-pooling embedding of ImageNet-pretrained ResNet-18, which never saw our labels. Features were standardised; embeddings were reduced to 50 principal components. We applied $k$-means and Gaussian mixture models (GMMs) with scikit-learn \citep{Pedregosa2011} and used labels only afterwards, to compute the adjusted Rand index \citep[ARI;][]{Hubert1985}, normalised mutual information (NMI) and purity; the silhouette score \citep{Rousseeuw1987} uses no labels. In this sample the purity of a single cluster is 0.558.

\begin{table}
\centering
\caption{Clustering of 4\,000 frames. Labels are used only to compute ARI, NMI and purity.}
\label{tab:cluster}
\small
\setlength{\tabcolsep}{3.5pt}
\begin{tabular}{@{}llcccc@{}}
\toprule
Features & Method & Sil. & ARI & NMI & Purity\\
\midrule
Hand-crafted & $k$-means, $k$=2 & 0.304 & 0.005 & 0.003 & 0.558\\
Hand-crafted & $k$-means, $k$=5 & 0.319 & 0.083 & 0.118 & 0.701\\
Hand-crafted & GMM, $k$=2 & -- & 0.000 & -- & 0.558\\
CNN embedding & $k$-means, $k$=2 & 0.174 & 0.488 & 0.391 & 0.850\\
CNN embedding & $k$-means, $k$=5 & 0.220 & 0.284 & 0.389 & 0.918\\
CNN embedding & GMM, $k$=2 & -- & \textbf{0.548} & -- & \textbf{0.870}\\
\bottomrule
\end{tabular}
\end{table}

\begin{figure*}
\centering
\includegraphics[width=\textwidth]{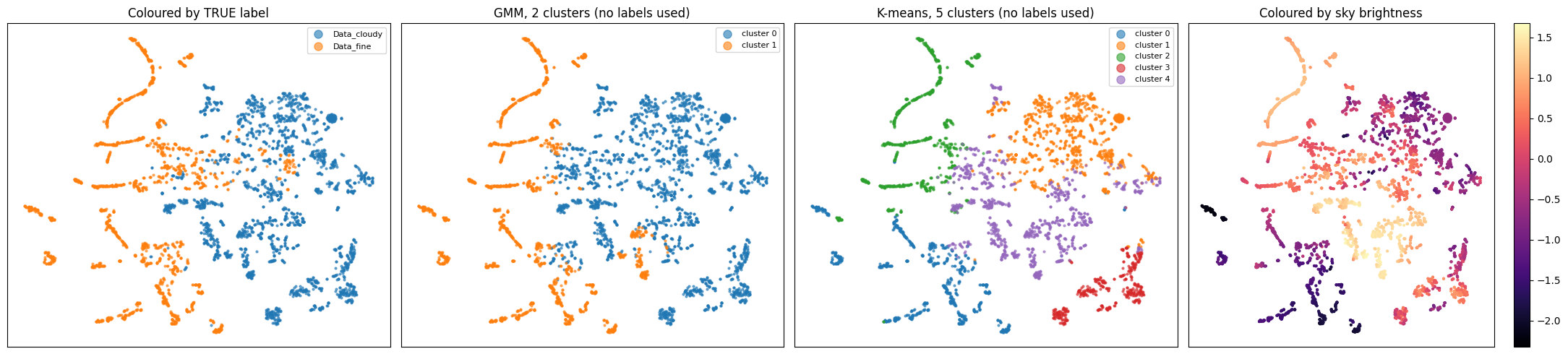}
\caption{t-SNE map of the pretrained ResNet-18 embeddings of 4\,000 frames, coloured (left to right) by dataset label, by the two GMM clusters (no labels used), by the five $k$-means clusters and by mean sky brightness.}
\label{fig:tsne}
\end{figure*}

The two descriptions behave very differently (Table~\ref{tab:cluster}). Hand-crafted statistics form well-separated clusters (silhouette $\approx$0.3) that are unrelated to the labels (ARI 0.005 for $k=2$): the two clusters have mean standardised brightness of $+0.83$ and $-0.88$, i.e.\ they split bright from dark skies, so the dominant variation in simple statistics is moonlight. Pretrained CNN embeddings instead recover the labels: a two-component GMM agrees with fine/cloudy for 87 per cent of the frames (ARI 0.548), comparable to the supervised test accuracy although on a different sample without held-out nights. In a t-SNE projection \citep{vanderMaaten2008} (Fig.~\ref{fig:tsne}) fine frames form continuous strands (clear nights evolving smoothly) and cloudy frames scattered clumps, and brightness runs across both classes: in CNN feature space, brightness and cloudiness are largely separate directions.

\begin{figure}
\centering
\includegraphics[width=\columnwidth]{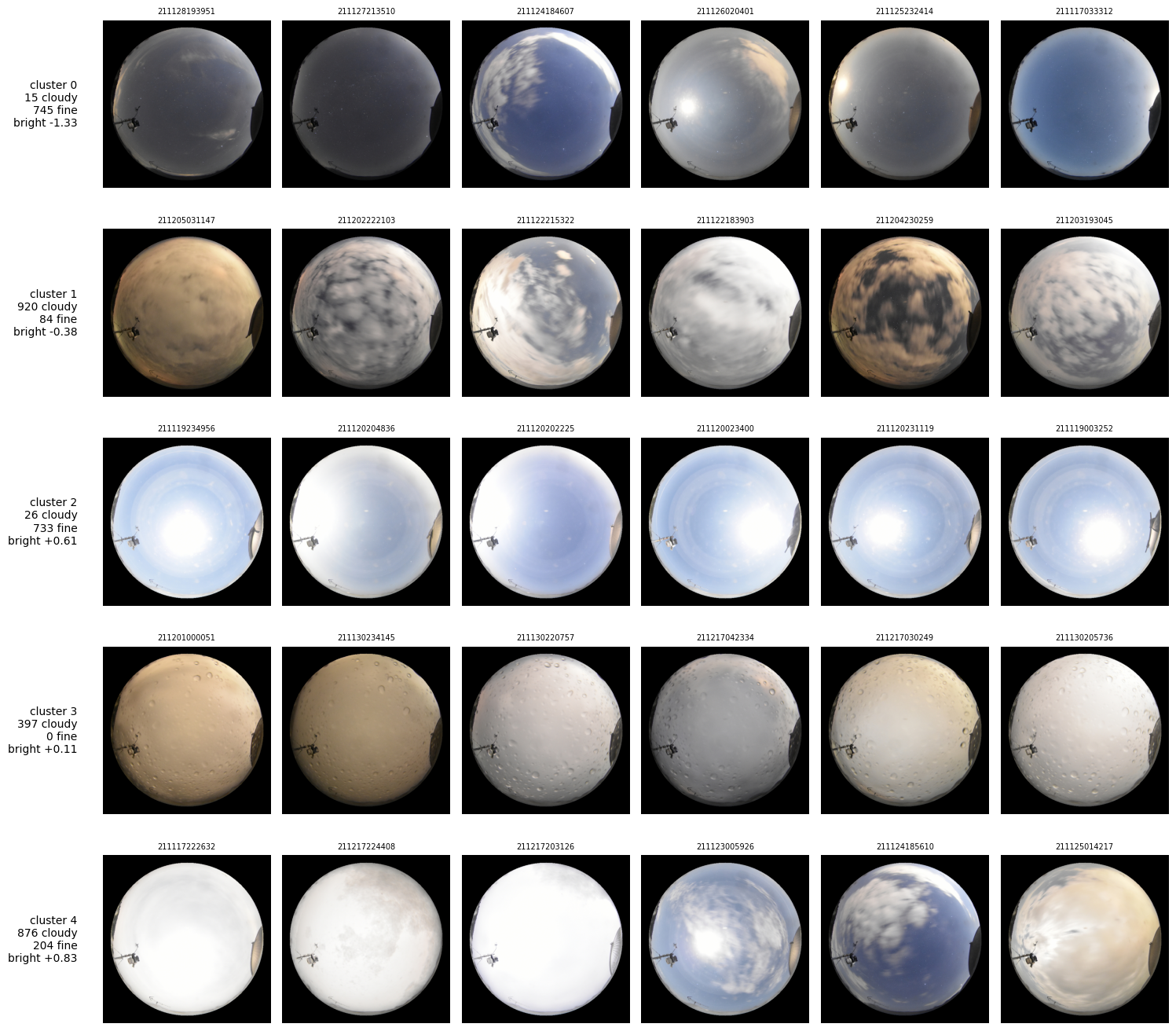}
\caption{Six random frames from each of the five $k$-means clusters with their composition. Cluster 3 consists of frames with water droplets on the camera dome.}
\label{fig:gallery}
\end{figure}

With $k=5$, $k$-means yields physically interpretable sky types (Fig.~\ref{fig:gallery}): dark clear skies (98 per cent fine; 15 cloudy and 745 fine frames), broken or textured cloud, often lit orange from below (92 per cent cloudy; 920/84), moonlit clear skies from the nights around the full Moon (97 per cent fine; 26/733), bright moonlit overcast and broken cloud (81 per cent cloudy; 876/204), and a cluster of 397 frames, all labelled cloudy, whose frames are covered with small round spots: \emph{water droplets on the camera dome}, from three rainy or humid nights. The labels merged this class with ordinary cloud, although operationally it calls for closing the dome rather than merely not observing. In practice, an operator could label a new dataset by naming a handful of clusters and reviewing only the ambiguous moonlit-cloud cluster frame by frame.

\section{Interpretation}
\label{sec:interp}

\begin{figure*}
\centering
\includegraphics[width=0.92\textwidth]{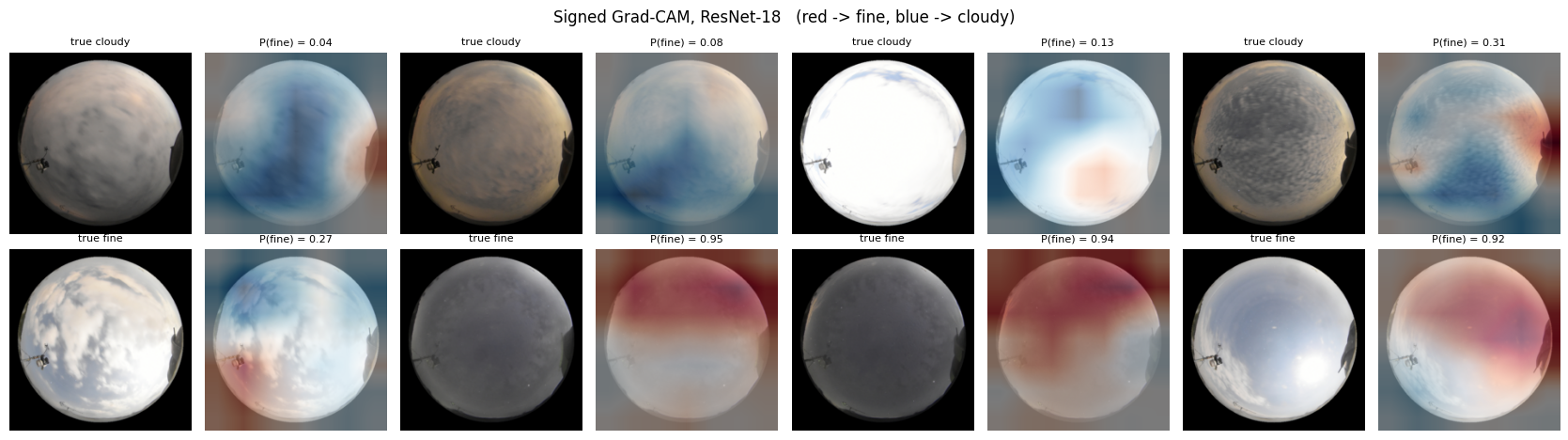}
\caption{Signed Grad-CAM of ResNet-18 (red: evidence for clear; blue: evidence for cloudy) for overcast, moonlit-overcast, broken-cloud, dark starry, moonlit clear and mislabelled frames.}
\label{fig:cam}
\end{figure*}

\begin{figure*}
\centering
\includegraphics[width=0.92\textwidth]{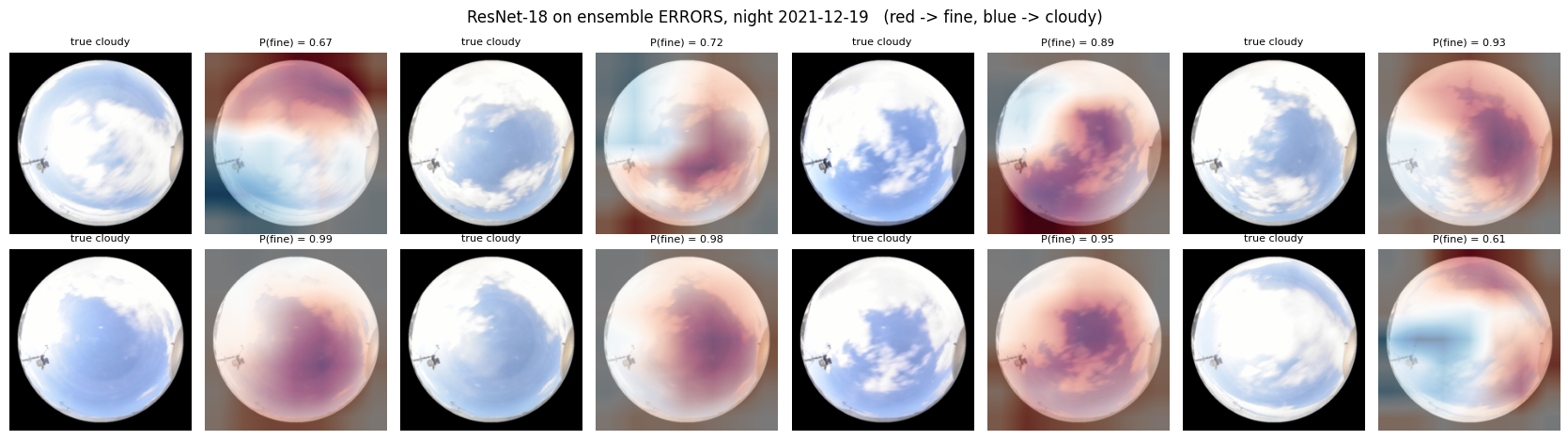}
\caption{Signed Grad-CAM of ResNet-18 for ensemble errors on the full-Moon night of 2021 December 19 (labelled cloudy, predicted clear). The evidence for clear lies on the clear patches of sky, not on the Moon's glare.}
\label{fig:cam1219}
\end{figure*}

\paragraph{Signed Grad-CAM.} Standard Grad-CAM \citep{Selvaraju2017} keeps only positive evidence for the predicted class; for cloudy frames the small CNN's maps were blank, suggesting that the network decides ``cloudy'' when it does not find clear-sky features. We therefore also computed a signed map, $L=\sum_k\alpha_k A^k$ with $\alpha_k$ the spatially averaged gradient of $z_{\rm fine}-z_{\rm cloudy}$ with respect to feature map $A^k$, without the final rectification. In the ResNet-18 maps (Fig.~\ref{fig:cam}) cloud-covered sky pushes towards cloudy, while star fields in dark frames and the blue sky around the Moon push towards clear; the Moon's glare itself does not. On the full-Moon errors of December 19, the evidence for clear lies on the clear gaps rather than on the glare (Fig.~\ref{fig:cam1219}), consistent with these frames being partly cloudy.

\paragraph{Rim test.} Because heat maps of a few frames suggested a hot spot at the edge of the sky circle, we measured, for 200 random test frames, the share of the signed map's total absolute value lying in the outer 15 per cent of the sky radius, which covers 28 per cent of the sky area. The share is 0.30 for the small CNN and 0.24 for ResNet-18, i.e.\ in proportion to or below the area: there is no systematic reliance on the rim, the mask edge or the fixed structures near it.

\paragraph{Occlusion.} Grey occluders resemble cloud on a night sky, so we replaced each $32\times32$ patch (stride 16) by a strongly blurred copy of itself (Gaussian $\sigma=12$ pixels), which preserves brightness and colour but removes stars and texture \citep[cf.][]{Zeiler2014}. Confident predictions are holistic: for frames with $P\approx0.99$--1.00, blurring any single patch changes $P$ by at most $6\times10^{-4}$ to $6\times10^{-3}$. Sensitivity appears only for uncertain frames, up to $\pm0.35$, and mainly along cloud edges. The strongest responses occur where patches straddle the mask boundary and are artefacts of the method.

\section{Exact evidence maps and the Sky Monitor AI application}
\label{sec:tool}

All three classifiers end with global average pooling followed by one linear layer. Let $\mathbf{A}(x,y)$ be the feature vector at position $(x,y)$ of the final feature map ($14\times14\times256$ for the small CNN, $7\times7\times512$ for ResNet-18), $N$ the number of positions, and $\mathbf{W},\mathbf{b}$ the weights and biases of the linear layer. Because averaging and the linear layer commute,
\begin{equation}
z_{\rm fine}-z_{\rm cloudy} = \frac{1}{N}\sum_{x,y}\Big[\underbrace{\Delta\mathbf{w}\cdot\mathbf{A}(x,y)+\Delta b}_{\ell(x,y)}\Big],
\end{equation}
with $\Delta\mathbf{w}=\mathbf{W}_{\rm fine}-\mathbf{W}_{\rm cloudy}$ and $\Delta b=b_{\rm fine}-b_{\rm cloudy}$. The local logit $\ell(x,y)$ is the decision the network would make if the whole image looked like position $(x,y)$, and the actual decision is exactly the mean of the local decisions. This is the class activation mapping of \citet{Zhou2016} applied to the logit difference, with the bias distributed over positions; unlike gradient-based maps it requires no gradients and is exact at the final layer. The sigmoid of $\ell$ gives a local probability of a clear sky. Upsampling the three maps to $224\times224$ and averaging them gives an ensemble evidence map, and we define the estimated cloud cover as the fraction of sky pixels with local probability below 0.5. In tests the mean of each local-logit map agreed with the model's logit difference to within $10^{-5}$.

Sky Monitor AI is a Streamlit application that runs the three classifiers and the embedding network, exported to ONNX and executed with ONNX Runtime on a CPU; the exported models reproduce the PyTorch outputs to within $4.4\times10^{-6}$ and the test-set probabilities exactly. For each frame it reports $P_{\rm ens}$, the evidence map, the cloud cover, the nearest of the five sky types, and an out-of-distribution warning when the embedding lies more than 1.2 times the 99th-percentile within-cluster distance from its nearest centre. A rule-based layer maps these quantities to five operational conditions: \emph{clear} ($P\geq0.7$ and cover $\leq0.3$), \emph{overcast} ($P\leq0.3$ and cover $\geq0.7$), \emph{mostly clear} or \emph{mostly cloudy} in between according to $c=\tfrac12(1-P)+\tfrac12\,{\rm cover}$, and \emph{wet dome} whenever the nearest sky type is the droplet cluster. These intermediate and wet-dome conditions are derived quantities, not separately trained or validated classes. The application analyses single frames, whole nights (with Moon altitude and trailing-window smoothing for real-time use), shared folders, and live cameras (a folder written by camera software, a latest-image URL, or a USB or RTSP stream), and can display a weather forecast and a geostationary satellite image of the site for comparison. Code and documentation are public (see Data availability).

\section{Discussion}
\label{sec:discussion}

\paragraph{Task difficulty and model capacity.} All frames come from one camera with fixed settings and one field of view, which makes the classification task comparatively simple and helps explain why every model reached reasonable performance. Even so, the choice of model capacity relative to the available data mattered. The effective sample size is of order the number of nights ($\sim$24 for training), not the number of frames; letting more of ResNet-18 adapt consistently increased the validation loss, and the best ResNet was a linear classifier on frozen features. The 389\,890-parameter small CNN was the best single model on the test nights, and several models showed large differences between validation and test performance (e.g.\ v2: 0.906 versus 0.812 accuracy). These observations do not by themselves establish \emph{why} the models differ, but they emphasise that generalisation must be measured on nights not used for model selection, and that large overparameterised models, although often successful, do not make capacity irrelevant for generalisation or for computational efficiency. In this respect our results are consistent with \citet{Mommert2020}, who found that gradient-boosted trees on informative features outperformed a ResNet on their data; here, with whole-night evaluation, CNN features outperform our hand-crafted features, mainly because they capture spatial structure and are less sensitive to moonlight.

\paragraph{Labels.} The block labelling limits both training and evaluation. A substantial fraction of the remaining errors are mislabelled or partly cloudy frames, so frame-level labels, or a continuous cloud-fraction target, would likely improve both the models and the reliability of the measured accuracy. The wet-dome cluster suggests that a third class, precipitation or condensation on the optics, is operationally more important than a finer subdivision of cloud cover \citep[cf.][]{Haghighi2025}.

\paragraph{Limitations.} The data span 32 nights from a single season and camera; the models have not been tested on other cameras, other seasons or in daytime, and the out-of-distribution warning is a simple heuristic. The analysis of Moon altitude rests on six test nights. The tool has not been operated alongside human observers, and should not control equipment unattended. Natural extensions are more nights with varied Moon phases, frame-level relabelling around weather transitions, sequence models that use cloud motion \citep[cf.][]{Buntin2025}, and comparison with the model used operationally at the observatory \citep{Uemura2025}.

\section{Conclusions}
\label{sec:conclusions}

We have presented an end-to-end study of night-sky classification for the all-sky camera of the Higashi-Hiroshima Observatory. A sky mask removes a date-stamp shortcut; on six held-out nights a validation-selected ensemble of three CNNs reaches 86.7 per cent accuracy and AUC 0.924 with the smallest night-to-night scatter, compared with 77.1 per cent for hand-crafted sky statistics. Moonlight is the main difficulty, and the ensemble is the least sensitive to it. Fine-tuning deeper layers overfits with $\sim$24 training nights. Many remaining errors reflect block labelling and partly cloudy skies. Pretrained CNN embeddings recover the clear/cloudy split without labels and reveal a wet-dome class, and interpretation maps indicate that the networks base their decisions on sky regions. An exact local-logit decomposition provides evidence maps and cloud-cover estimates in an open-source application that can be used with archived frames and live cameras.

\section*{Acknowledgements}
The all-sky camera images were acquired at the Higashi-Hiroshima Observatory, Hiroshima University, and were provided by Prof.\ Makoto Uemura to the AstroAI Network Summer School 2026 (hosted by ASIAA and LeCosPA, National Taiwan University) as a participant project. I thank Prof.\ Uemura for making the data available and for helpful comments on the work, and the organisers of the summer school. The author is supported by a Kemitraan Negara Berkembang (KNB) scholarship of the Government of Indonesia. This work made use of PyTorch, scikit-learn, Astropy, ONNX Runtime and Streamlit, and of Google Colab computing resources.

\section*{Data availability}
The code, the application and the trained models are available at \url{https://github.com/waqas6767/Sky-Monitor-AI-for-Remote-Observatory-Operations} (models in release v1.0); a live demonstration is available at \url{https://sky-monitor-ai-for-remote-observatory-operations-ybn86wj9z54ze.streamlit.app/}. The all-sky camera images are not part of this release; requests should be directed to the Higashi-Hiroshima Observatory, Hiroshima University.

\bibliographystyle{plainnat}
\bibliography{refs}

\end{document}